\documentclass[fleqn,usenatbib]{rasti}

\usepackage{newtxtext,newtxmath}

\usepackage[T1]{fontenc}

\DeclareRobustCommand{\VAN}[3]{#2}
\let\VANthebibliography\thebibliography
\def\thebibliography{\DeclareRobustCommand{\VAN}[3]{##3}\VANthebibliography}

\usepackage{graphicx}	
\usepackage{amsmath}	

\graphicspath{{figures/}}

\newcommand{\urlVC}{https://github.com/rgbIAA/viewcube}

\newcommand{\urlVCeval}{https://github.com/rgbIAA/ViewCube-Evaluation/tree/main/2024}

\title[Interactive Multimodal IFS]{Interactive Multimodal Integral Field Spectroscopy}

\author[A. García Riber et al.]{
Adrián García Riber,$^{1}$
Rubén García-Benito,$^{2}$\thanks{E-mail: rgb@iaa.es}
and Francisco Serradilla$^{3}$
\\
$^{1}$DMCT, ETSIST, Universidad Politécnica de Madrid (UPM), 28031, Spain\\
$^{2}$Instituto de Astrofísica de Andalucía - CSIC, Glorieta de la Astronomía s.n., 18008 Granada, Spain\\
$^{3}$Departamento de Sistemas Informáticos, ETSISI, Universidad Politécnica de Madrid (UPM), 28031, Spain
}

\date{Accepted REVIEWED. Received YYY; in original form ZZZ}

\pubyear{2024}

\begin{document}
\label{firstpage}
\pagerange{\pageref{firstpage}--\pageref{lastpage}}
\maketitle

\begin{abstract}
Using sonification on scientific data analysis provides additional dimensions to visualization, potentially increasing researchers’ analytical capabilities and fostering inclusion and accessibility. This research explores the potential of multimodal Integral Field Spectroscopy (IFS) applied to galaxy analysis through the development and evaluation of a tool that complements the visualization of datacubes with sound. The proposed application, \textit{ViewCube}, provides interactive visualizations and sonifications of spectral information across a two-dimensional field-of-view, and its architecture is designed to incorporate future sonification approaches. The first sonification implementation described in this article uses a deep learning module to generate binaural unsupervised auditory representations. The work includes a qualitative and quantitative user study based on an online questionnaire, aimed at both specialized and non-specialized participants, focusing on the case study of datacubes of galaxies from the Calar Alto Integral Field Spectroscopy Area (CALIFA) survey. Out of 67 participants who completed the questionnaire, 42 had the opportunity to test the application in person prior to filling out the online survey. 81\% of these 42 participants expressed the good interactive response of the tool, 79.1\% of the complete sample found the application "Useful", and 58.2\% rated its aesthetics as "Good". The quantitative results suggest that all participants were able to retrieve information from the sonifications, pointing to previous experience in the analysis of sound events as more helpful than previous knowledge of the data for the proposed tasks, and highlighting the importance of training and attention to detail for the understanding of complex auditory information.

\end{abstract}

\begin{keywords}
Software -- Data Methods -- Spectroscopy -- Sonification -- Machine Learning
\end{keywords}


\section{Introduction}
The combination of visual and auditory displays can offer a better understanding of a phenomenon~\citep{enge2024open}, making the use of sound for the representation of physical quantities an established area of research \citep{dubus2013systematic}. It can allow holistic interpretations of the data facilitating the discovery of previously unseen relationships~\citep{cooke2017exploring}, as well as single datum analytic tasks, both involving point estimation and comparison, trend identification, and data structure analysis~\citep{walker2011theory}.  

Sonification has proven to be a very useful tool to assist in the analysis of hyperspectral data, generating sonic time-series related to the spatial and spectral content near user-selected mouse positions \citep{Bernhardt:2007}. It can also improve the perception of density in visualization of complex data in parallel coordinates and scatter plots \citep{ronnberg2016interactive}, and ensure simple access to information for blind and non-blind people, enhancing the accessibility of astronomical data as well as the work of the scientist accepting complementary exploration methodologies \citep{casado2024analysis}.

Furthermore, the enhancement of visual information with sonification allows sighted and blind or low vision (BLV) communities to have astronomy experiences at similar levels~\citep{arcand2024universe}. Including blind users from the beginning of the design process, \cite{casado2024multimodal} proposed the case study of the galaxies from the \textit{Sloan Digital Sky Survey} (SDSS) to show the possibilities of \textit{sonoUno}. This multimodal application for displaying sound and images from any dataset allowed for the discovery of the variable star UCAC4 459-09273 by blind students using sonification. As a blind researcher, \cite{foran2022power} reported the use of \textit{StarSound} on 1D high-redshift galaxy work for the verification and initial analysis of the rest-frame ultraviolet (UV) spectra of distant galaxies, and developed the touch-based sonification tool \textit{VoxMagellan} to analyze 2D images and multidimensional data sets. 


In the photometric and spectroscopic analysis field, \cite{trayford2023inspecting} proposed the audification of spectral datacubes (direct conversion of data into audible frequencies), to demonstrate that physical information can be extracted directly from sound with STRASUSS~\citep{trayford2023introducing}. Using \textit{Star Sounder}, ~\cite{huppenkothen2023sonified} provided an interactive sonification of the Hertzsprung-Russell diagram based on the crossmatch between the Kepler Stellar Table and Gaia Data Release (DR2). The introduction of a sonic perspective has also been valued on spectroscopic analysis of Quasars by ~\cite{hansen2020quasar}, finding that sonification can enable more rapid discovery and identification of intergalactic/circumgalactic medium (IGM/CGM) system candidates than visually scanning through spectra.

Including multimodal interactivity, \cite{Starks} explored the data-cubes of the \textit{Antennae Galaxies} radio-image from the Atacama Large Milimeter/submilimeter Array (ALMA), using \textit{Galaxy player} within the \textit{Soniverse} project. Additionally, spatialized sonifications have been used on immersive representations of Antarctic astronomy data~\citep{west2018experiencing}, and highlighted by~\cite{quinton2021sonification, quinton2020sonification} as an effective parameter mapping strategy with the potential to detect sudden changes between multiple sources within the field of exosolar planetary search.
 
Framed in this context of inclusive and immersive scientific representations, this article presents \textit{ViewCube}, a multimodal interactive binaural tool for the analysis of datacubes with headphones. The application includes an unsupervised sonification approach, based on autoencoders. Furthermore, the work explores the potential of multimodal Integral Field Spectroscopy (IFS) using the case study of the Calar Alto Legacy Integral Field Spectroscopy Area (CALIFA) survey \citep{Sanchez:2012, sanchez2016califa}.

Providing quantitative and qualitative feedback from specialized and non-specialized users in Astronomy and Music, this paper aims at demonstrating the usefulness of sound in multimodal displays on IFS analysis to:
a) estimate the position of the user-selected spaxel within the galaxy, identifying whether it is located to the left/right or front/rear in the virtual soundscape; b) estimate the spaxel's relative distance from the center of the galaxy, indicating whether it is near or far from this reference point; and c) identify the type/age of a spectrum, determining whether the spaxel corresponds to a star-forming region (spectrum with multiple and relatively strong emission lines), to an intermediate age galaxy, or to a retired galaxy. The work is expected to make IFS more accessible for BLV researchers while also enhancing the overall capabilities for datacube analysis.



\section{Multimodal IFS}

This section describes the design strategy and implementation of \textit{ViewCube}. The application combines graphical and auditory displays for the interactive multimodal analysis of datacubes. Using the case study of the CALIFA survey datacubes, the work aims at enhancing 3D spectroscopy representation with immersive sonification.

\subsection{Case study}
The CALIFA \citep{sanchez2016califa, sanchez2012califa, walcher2014califa} is a public legacy survey of over 600 galaxies  with an r-band isophotal major axis between 45" and 79.2" and a redshift 0.005<z<0.03, selected from the Sloan Digital Sky Survey (SDSS) DR7 photometric catalog. Aimed at helping in the study of galaxy evolution in the Local Universe through cosmic time, it uses integral field spectroscopy \citep[IFS,][]{allington2006basic} to provide a wide-field IFU survey of galaxies that includes all morphological types, covering masses between 10\textsuperscript{8.5} and 10\textsuperscript{11.5} M\textsubscript{$\odot$} \citep{sanchez2016califa}.


The observations were obtained with the integral-field spectrograph PMAS/PPak mounted on the 3.5 m telescope at the Calar Alto observatory. The wavelength range between 3700 and 7500 \AA\ is sampled using two different spectral setups, a low resolution V500 mode (3745–7500 \AA) with a spectral resolution of 6.0 \AA\ (full width at half maximum, FWHM), and a medium-resolution V1200 mode (3650–4840 \AA) with a spectral resolution of 2.3 \AA\ ~\citep[FWHM,][]{garcia2015califa}. CALIFA's third Data Release ~\citep[DR3,][]{sanchez2016califa} provides to the public 646 objects in the V500 setup, 484 in the V1200, and the combination of the cubes from both setups (COMBO).

The morphology references provided for each galaxy used in this work are extracted from \cite{walcher2014califa}.

\subsection{Standalone application}

\textit{ViewCube}\footnote{The application can be downloaded from: \url{\urlVC}} is a lightweight, standalone application written entirely in Python, designed for the efficient browsing of datacubes. Originally developed for the quick assessment of the quality and physical characteristics of datacubes from the CALIFA survey, and for the rapid exploration of high-level data products generated by the PyCASSO pipeline \citep{deAmorim:2017}, \textit{ViewCube}'s functionality has since expanded. The application now supports datacubes from any provenance, thanks to its general and flexible FITS reader. This reader is agnostic to the source of the datacubes, enabling it to handle a wide variety of datacubes from different instruments (e.g., MUSE) and surveys (e.g., MaNGA, CALIFA) across various wavelength ranges (optical, radio).

Despite its name, \textit{ViewCube} is also capable of rendering Raw Stacked Spectra (RSS) formats, provided that a file mapping the positions of the fibers is available. Visualizations within \textit{ViewCube} are rendered using the Matplotlib module \citep{Hunter:2007}. The primary objective of \textit{ViewCube} is to facilitate a fast and effective inspection of datacubes, either for a quick quality assessment or for a focused examination of their characteristics.

\begin{figure}
	\includegraphics[width=\columnwidth]{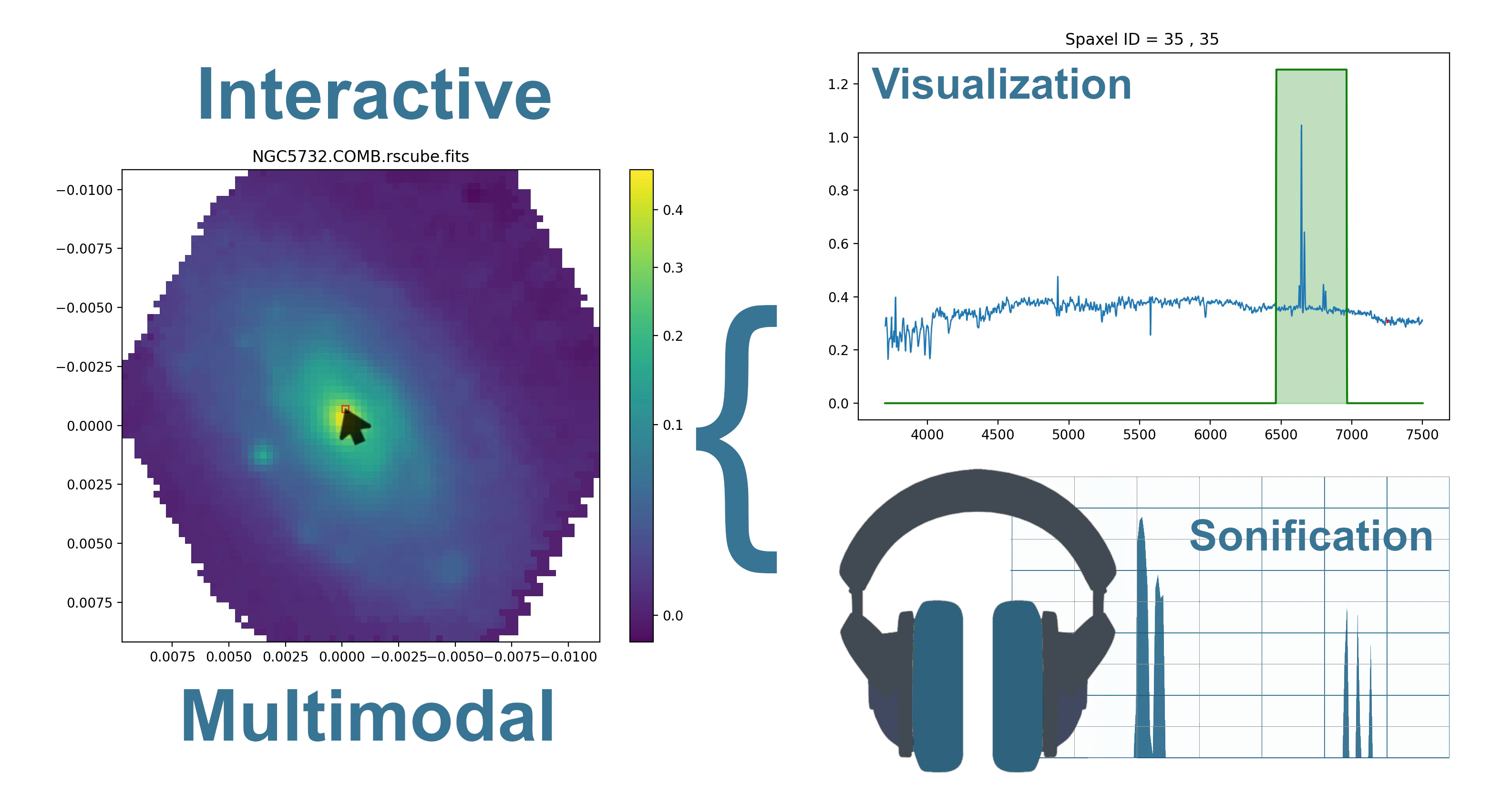}
    \caption{\textit{ViewCube} UI displaying the datacube of the spiral (Sbc) galaxy NGC 5732. 2D image window (left) and multimodal representation --spectral window and sonification-- of the spaxel (35,35) (right).}
    \label{fig:UI_figure}
\end{figure}

As shown in Fig.~\ref{fig:UI_figure}, the user interface features two main windows: an image window, which presents a 2D map of the datacube convolved through a chosen passband, and a spectral window, which displays the spectrum corresponding to the location of the mouse pointer. Users can select different spaxels or fibers for comparison, generate an integrated spectrum, and save both individual and integrated spectra. Additionally, users can modify the filter used to convolve the datacube to produce the image in the image window, and adjust the central wavelength of the filter by dragging and dropping the filter passband shown in the spectral window.

In keeping with its initial exploratory purpose for quality assessment, \textit{ViewCube} allows for the interactive comparison of spectra from two datacubes at the same spaxel, provided the datacubes share the same dimensions. To enhance the exploration of individual spectra and perform more advanced operations, \textit{ViewCube} offers the possibility of integrating with other packages such as PySpeckit \citep{Ginsburg:2022} and PyRAF \citep{pyraf}. Future versions of \textit{ViewCube} are planned to include a faster rendering visualization engine, as well as additional menu options for improved functionality. 

\begin{figure*}
	\includegraphics[width=140mm]{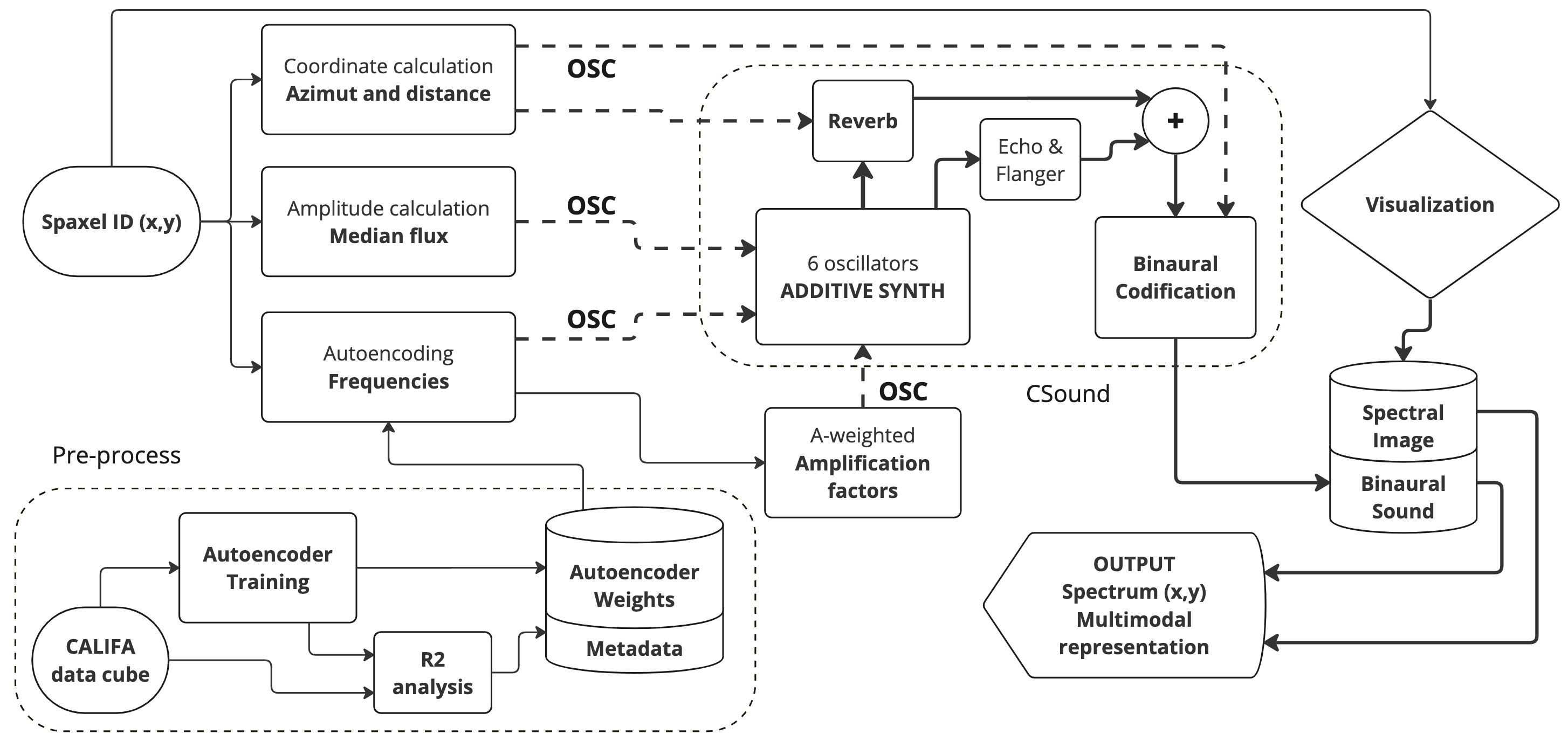}
    \caption{\textit{SoniCube} block diagram. Pre-processing and real time calculations including data, OSC, and audio signal flows.}
    \label{fig:Block_figure}
\end{figure*}

\subsection{Sonification module}
This section describes the sound module implemented within \textit{ViewCube} to allow the sonification of the spectra associated with each spatial element of a datacube. The module, named \textit{SoniCube}, provides an open, comprehensive, and general-purpose multimodal tool for IFS analysis, supporting the development of future sonification techniques focused on specific potential features within datacubes. The aim of the \textit{SoniCube} interface is to offer a diverse "palette of sonifications",  akin to the range of color palettes available for visual 2D maps in \textit{ViewCube}. This variety of sonification options allows users to extract or enhance different data characteristics by selecting specific sonification methods, much like how a color palette reveals visual details. In this paper, we present the first sonification method implemented within \textit{SoniCube}'s palette, providing an autonomous fast representation of the spectra. This first  sonification 'palette' implementation is designed to replicate the purpose of the visual \textit{ViewCube} counterpart, providing a quick qualitative overview of the datacube.

Also implemented in Python, \textit{SoniCube} controls a sound synthesizer developed in CSound via Open Sound Control (OSC)~\citep{ref_OSC}, using the python-osc native module and the ctcsound interface~\citep{ref_ctcsound}. The module provides a real-time interactive sonification of the spectrum associated to each user-selected spaxel. In this first sonification implementation, each spectrum is converted into sound using the deep learning approach described in Section~\ref{sec:DL_Sonification}. This process provides an unsupervised unique auditory footprint conveying the information of each spectrum in one single sound event. Each sound event is generated with an additive synthesizer using six independent oscillators, fed by a six-dimensional latent vector which is generated by an autoencoder \citep{baldi2012autoencoders}.

The module generates "on the fly" a six-dimensional latent vector from each user-selected spectrum. The components of this vector are interpreted as fundamental frequencies for the six oscillators that synthesize the sound. The six components are multiplied by a factor of 10,000 for scaling the latent values to audible frequencies, generating comprehensible accurate sonifications. For a formalized description of the synthesizer see Appendix~\ref{sec:apendix}.

Additionally, the module calculates the azimuth and radial distance from user-selected spaxels to the reference spaxel, which corresponds to the center of the galaxy on each datacube. The azimuth is used to locate the auditory footprint of the spectrum within the binaural soundscape~\citep{moller1992fundamentals} generated for each datacube, providing an immersive representation of its spectra with the listener located at the center of the galaxy. For more information about binaural encoding see Appendix~\ref{sec:apendix2}.

The distance is used to feed the direct-to-reverberant energy ratio of a reverberation emulator \citep{gardner1998reverberation}, which provides the cognitive sensation associated with the sound field that can be found in large indoor environments. This effect is used to generate a virtual auditory cue for distance perception based on direct-to-reverberant energy ratio~\citep{lu2010binaural}, related to the proximity from the user-selected spaxel to the reference spaxel in the center of the galaxy.

The amplitude of each sonification is calculated from the absolute fluxes of the represented spectrum. To solve the wide and variable dynamic range of flux density commonly found in real sky observations, \textit{SoniCube} includes two operating modes for representing relative sound amplitudes. In the flux sensitive mode (default), the amplitude of each sonification is calculated from the logarithmic median of the absolute flux, normalized using feature scaling. This mode preserves the apparent relation of fluxes within the datacube. On the other hand, if the sensitive mode is deactivated, all spectra in the datacube are represented with the same amplitude, allowing the appreciation of regions with relatively low absolute fluxes. An additional two-stage broadband dynamic range limiter/compressor \citep{kates2005principles} is implemented in both modes to keep extremely salient values controlled, preventing from hearing damage when analyzing unexplored datacubes.

To compensate for the non-linear response of human hearing in frequency and amplitude domains, an A-weighted curve is applied to the array of amplitudes using the librosa module~\citep{mcfee2015librosa}. This function calculates the normalized amplification factors that provide an equal loudness response on each footprint, presenting good results for the implementation of this first sonification module evaluated with the CALIFA survey. Nevertheless, alternative loudness contours are being explored for critical future implementations~\citep{charbonneau2012weighting}.

Finally, using the Open Sound Control (OSC) protocol, the module sends to CSound all the variables needed to synthesize the auditory footprint in a binaural soundscape. The block diagram of Fig.~\ref{fig:Block_figure} summarizes the sound generation process. Notice the use of azimuth and distance of the selected spaxel in the binaural and reverberation blocks, as well as the normalized median flux, and the frequencies with A-weighted factors of the corresponding spectrum in the additive synthesizer. 

Addressing some additional aesthetic aspects of the sonification, background echo and flanger effects were added to the workflow, facilitating the binaural localization of the footprint and smoothing fast transitions between spectra\footnote{The following video shows a demonstration of the application.

\url{https://vimeo.com/1005208084}}.

\subsection{Autoencoding CALIFA}
\label{sec:DL_Sonification}

As mentioned in the previous section, this first sonification implementation of \textit{SoniCube} uses an autoencoder architecture to provide accurate sonifications of the spectral information of a datacube. Based on the gradient descent algorithm, these networks allow the reduction of data dimensionality better than other approaches such as principal components analysis ~\citep[PCA,][]{hinton2006reducing}. Autoencoders are neural network models with the potential to learn an approximation to the identity function, providing an output that is similar to their input~\citep{ng2011sparse}. By reducing the number of hidden units of the intermediate layer of the network, a model can learn relevant structures of the data, which can also be reconstructed from this intermediate lower-dimensional representation, named latent space~\citep{goodfellow2016deep}. 

The dimension of the latent space depends on the data set and the architecture used. Aimed at obtaining stellar parameters using convolutional neural networks,~\cite{mas2024using} proposed a 32-dimensional latent space autoencoder, displayed in a 8x4 matrix, for the reduction of CARMENES spectra, and ACES synthetic spectra modeled with~\cite{phoenix1990}. On the other hand, the reconstruction from four-dimensional latent vectors was enough for \cite{xiang2022investigation} to analyze the stellar magnetic activity using variational autoencoders on the Large Sky Multi-Object Fiber Spectroscopic Telescope (LAMOST) K2 spectra. Demonstrating the potential of variational autoencoders,~\cite{portillo2020dimensionality} summarized galaxy spectral information with only six latent variables on the Sloan Digital Sky Survey (SDSS). This dimension also worked effectively for the Calcium II Triplet library (CaT) reduced with sparse autoencoders \citep{garcia2024deep}, which also agrees with our preliminary tests on the CALIFA survey galaxies.

\begin{figure}
	\includegraphics[width=\linewidth]{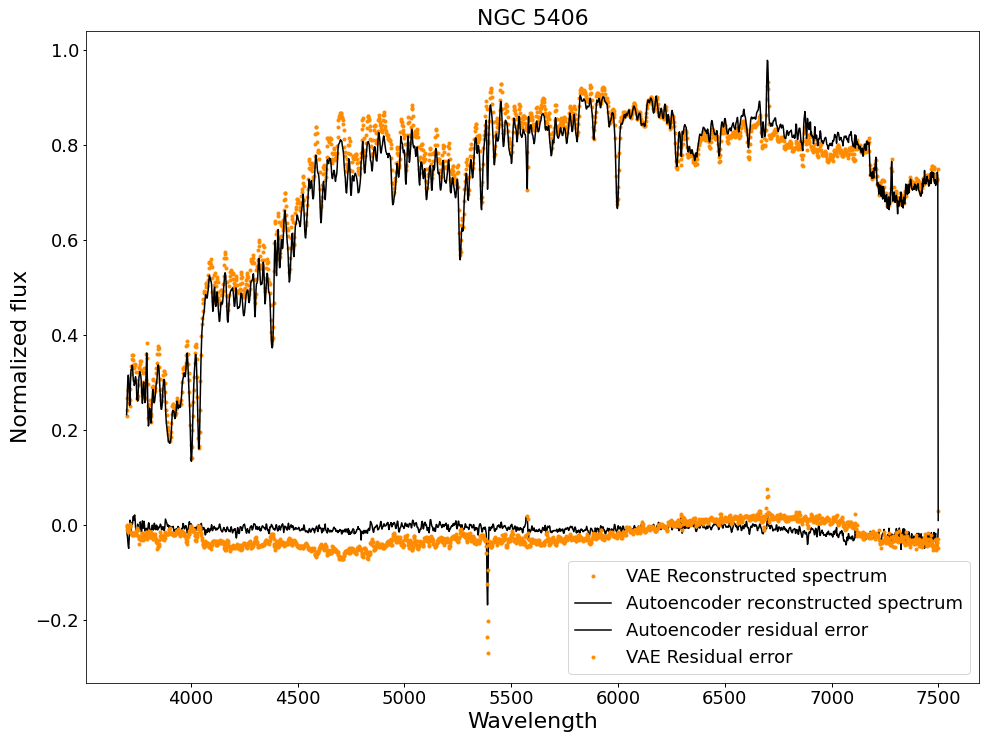}
    \caption{Autoencoder comparative for the spiral (Scd) galaxy NGC 5406. Six-layer six-dimensional autoencoder (black, continuous) vs four-layer six dimensional VAE (orange, dotted). Reconstructed spectra and residual error from the original spectrum for spaxel (34,34). Sparse autoencoder: R\textsuperscript{2} = 0.99 (spectrum), R\textsuperscript{2} = 0.98 (datacube), 39.12\% of the spectra with R\textsuperscript{2} > 0.9, 100 epochs, one hour per cube. VAE: R\textsuperscript{2} = 0.97 (spectrum), R\textsuperscript{2} = 0.98 (datacube), 4.92\% of the spectra with R\textsuperscript{2} > 0.9, 291 epochs, 5h 30' per cube. Normalized flux (ADU) vs wavelength (\r{A}).}
    \label{fig:Comparison_figure}
\end{figure}

Intensive testing was done with different configurations of both architectures using the COMBO (V500+V1200) datacubes of the DR3 CALIFA survey. Fig.~\ref{fig:Comparison_figure} provides a comparative example for the datacube of the spiral (Scd) galaxy NGC 5406 with a six-layer six-dimensional autoencoder, and a four-layer six-dimensional VAE, both implemented using TensorFlow~\citep{abadi2016tensorflow}. 

On each encoded datacube, we calculated the coefficient of determination (R\textsuperscript{2}) between the original and the reconstructed sets of spectra, providing a measure of the accuracy of the reduction. As for the duration of their training processes, the VAE required 5.5 times more computation time per epoch than the sparse autoencoder to provide lower results. Respectively, R\textsuperscript{2} = 0.96 (VAE) vs R\textsuperscript{2} = 0.98 (sparse) for the complete datacube, with 4.92\% (VAE) vs 39.12\% (sparse) of the spectra with R\textsuperscript{2} > 0.9,  and R\textsuperscript{2} = 0.96 vs R\textsuperscript{2} = 0.99 for the represented spaxel (34,34). 

Based on these tests, a six-layer six-dimensional sparse autoencoder module was included in \textit{SoniCube} to represent "in real time" the spectral information of the datacubes with low-dimensional vectors. This architecture allowed the reduction of each input spectrum X\textsubscript{i} $\in$ $\mathbb{R}$\textsuperscript{1901} (each datacube contains around 5540 spectra with 1901 flux values per spectrum), to a six-dimensional representation Z\textsubscript{i} $\in$ $\mathbb{R}$\textsuperscript{6}, and the reconstruction of X\textsubscript{i} from the latent vector Z\textsubscript{i}, $\hat{X}$\textsubscript{i} $\in$ $\mathbb{R}$\textsuperscript{1901}.

\begin{figure*}
	\includegraphics[width=\linewidth]{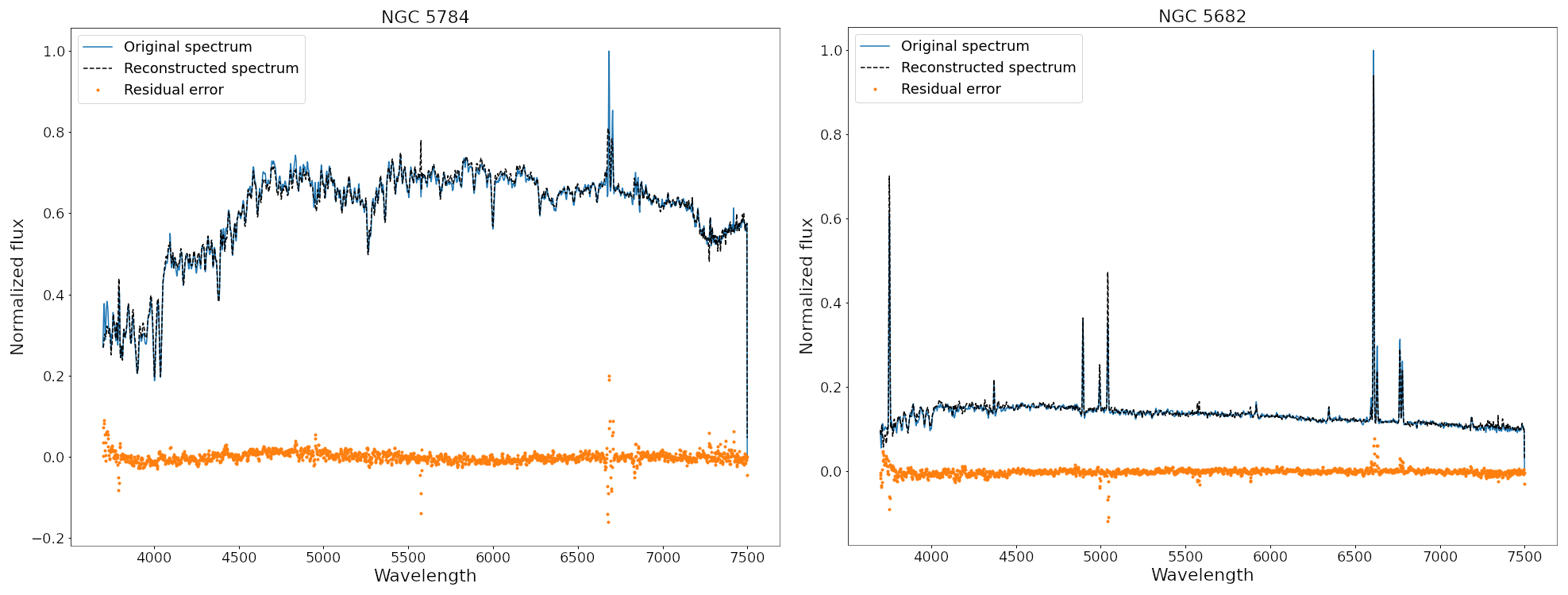}
    \caption{Six-layer six-dimensional autoencoder results. Reconstructed (black dashed line) and original (blue solid line) spectrum with residual error (orange dotted line) of the spaxel (35,35) from the spiral (Sbc) galaxy NGC 5784 (left), and the elliptical (E4) galaxy NGC 5682 (right). Two examples of an old galaxy and a star-forming region from the CALIFA survey. Respectively, R\textsuperscript{2} = 0.98 and R\textsuperscript{2} = 0.95. Normalized flux (ADU) vs wavelength (\r{A}).}
    \label{fig:Spectra_figure}
\end{figure*}

\begin{figure*}
	\includegraphics[width=\linewidth]{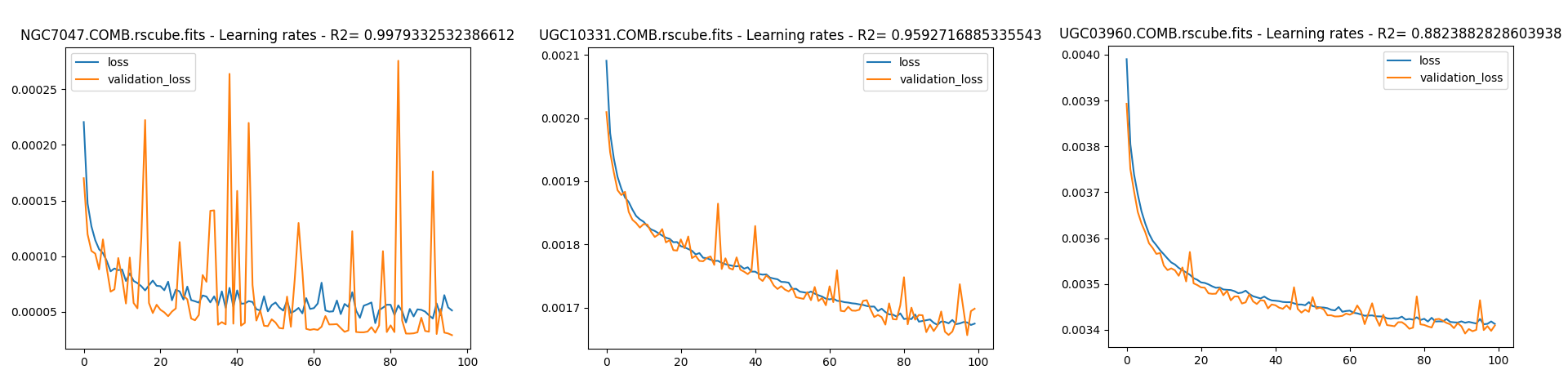}
    \caption{Learning curves showing mean square error vs epoch during the training and validation processes for the spiral (SAb) galaxy NGC7047, the elliptical (E1) galaxy UGC10331, and the elliptical (E5) galaxy UGC03960. These datacubes correspond respectively to the best, medium, and worst encoding results provided by the autoencoder. Notice the difference of scale in the y axis.}
    \label{fig:Learning_figure}
\end{figure*}

The model was trained on each datacube independently with around 5540 spectra. Fig.~\ref{fig:Spectra_figure} provides two examples of the original and reconstructed spectra from the datacubes of NGC 5784 (Sbc) and NGC 5682 (E4)\footnote{
The corresponding sonicubes, the encoded files for 446 galaxies from the DR3 COMBO datacubes of the CALIFA survey, can be downloaded from:
\url{https://zenodo.org/records/10570065}}.

Fig.~\ref{fig:Learning_figure} shows three examples of the learning curves obtained during the training process with the spectra from the datacubes of NGC7047 (Sab), UGC10331 (E1), and UGC03960 (E5). These galaxies illustrate the performance of the autoencoder, respectively corresponding to the best, the medium, and the worst encoding results. The coefficients obtained ranged from 0.998 to 0.882 along the complete data set, with 49\% of the datacubes presenting an R\textsuperscript{2} higher than 0.96, and 4.78\% presenting an R\textsuperscript{2} under 0.92.

\section{Evaluation}
To evaluate the potential utility of the previously described approach for the auditory analysis of galaxy datacubes, specifically using CALIFA datacubes as representative data, we conducted an anonymous online survey provided here for reference\footnote{The survey form is available at:

\url{https://forms.office.com/e/RbHVp8Vbbt}}.
The questionnaire was complemented by training videos and, to a lesser extent, in-person interactive demonstrations, targeting both specialized and non-specialized participants. All participants received the same online form, where they indicated whether or not they had experienced the application in person.

\begin{figure*}
	\includegraphics[width=\linewidth]{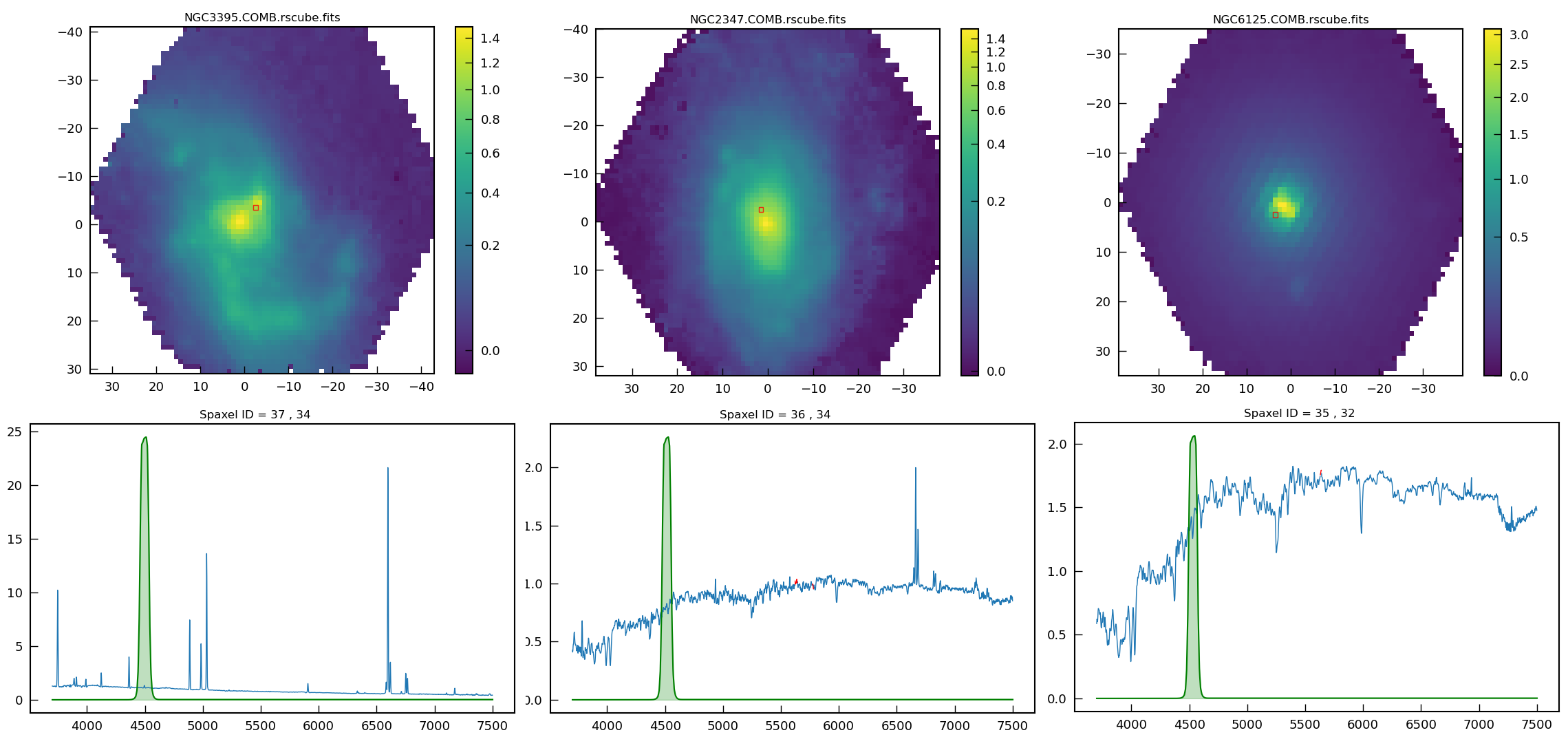}
    \caption{Age/Galaxy type examples presented in the training videos. Spectrum of a star-forming region in the spiral galaxy (S0) NGC 3395 (left), spectrum close to the center of the intermediate-age spiral galaxy (Sd) NGC 2347 (center), and spectrum of a region close to the center of the retired spiral galaxy (Sb) NGC 6125 (right). The upper panels display the narrowband image, produced using the narrowband filter indicated by the green curves in the lower spectral panels. The spectra corresponding to specific spaxels, highlighted by red squares, are indicated in the upper continuum maps. The axes of the upper panels represent offsets in arcseconds relative to the center of the galaxy. The spectra in the lower panels are plotted with wavelengths in Angstroms. The colorbar represents the flux of the data cube convolved with the filter, in units of 10$^{-16}$ erg cm$^{-2}$ s$^{-1}$.}
    \label{fig:Galaxies}
\end{figure*}

\subsection{Survey design}
\label{sec:Survey_design}

The online survey was administered to volunteer participants from April 15 to July 31, 2024. The questionnaire featured five training videos, which could be replayed as needed, including one providing a general overview of the application and one for each specific section. The survey comprised four sections with video-supported questions that analyzed various aspects of the proposal. Additionally, 12 questions were included to gather demographic information, participants' self-reported levels of expertise in Astronomy and Music, and three qualitative assessments concerning the application's interactivity, usefulness, and the aesthetics of the sounds employed in the sonification.

All participants were advised to use headphones and to check their correct placement in left and right ears. There was no time limit to complete the survey. The following describes the sections and questions included in the survey. Each question included several sonifications with no graphics, generated from the spectra of the galaxies NGC 5784 (Sbc), NGC 5732 (Sbc), NGC 5682 (E4), NGC 6060 (S0a), NGC 7562 (Sbc), NGC 7671 (S0), NGC7800 (Ir), NGC 2638 (Sb), and UGC 00148 (Sb). The participants could compare the questions with the examples presented in the training videos as many times as needed. 

\textbf{Section 1. Sound Location.} This first section consisted of four questions designed to analyze the possibilities of the application to estimate sound location. Within the virtual binaural soundscape provided, the listener is virtually placed in the center of the galaxy, looking at the upper position. The training videos of this section provided examples around the spiral galaxy (Sa) NGC 7549.

\textbf{Section 2. Distance to the center of the galaxy.} This section also included four questions aimed at studying the possibilities of the application to provide auditory information about the distance from the moving cursor to the center of the galaxy within the virtual spectral soundscape The training videos of this section provided examples from the spiral galaxy (Sbc) NGC 5732.

\textbf{Section 3. Age/Galaxy type.} 

This section explored how multimodal representation could aid in differentiating between various Age/Galaxy types. The training videos presented three examples, illustrated in Fig.\ref{fig:Galaxies}. These examples include: the spectrum (37,34) from a star-forming region of the spiral galaxy (S0) NGC 3395; the spectrum (36,34) near the center of the intermediate-age spiral galaxy (Sd) NGC 2347; and the spectrum (35,32) from a region near the center of the retired spiral galaxy (Sb) NGC 6125.

\textbf{Section 4. Combined questions.} Finally, two multiple choice questions analyzed the potential of the application to allow the identification of the position of the represented spectrum (left/right), its distance to the center (close/far), and if it corresponded to a star-forming region or to a retired galaxy.

\textbf{Qualitative question 1.} If you have tried the application in person, please rate the multimodal experience. If not (you only saw the training videos of this questionnaire), please skip this question.
Options: Very bad, Bad, Acceptable, Good, or Very good.

\textbf{Qualitative question 2.} Rate the potential usefulness of the multimodal display for the exploration of the CALIFA Survey. 
Options: Useless, Doubtfully useful, Useful, or Very useful.

\textbf{Qualitative question 3.} Rate the aesthetics of the sonifications. 
Options: Intolerable, Bad, Acceptable, Good, or Nice Sounding.

\begin{figure*}
	\includegraphics[width=\linewidth]{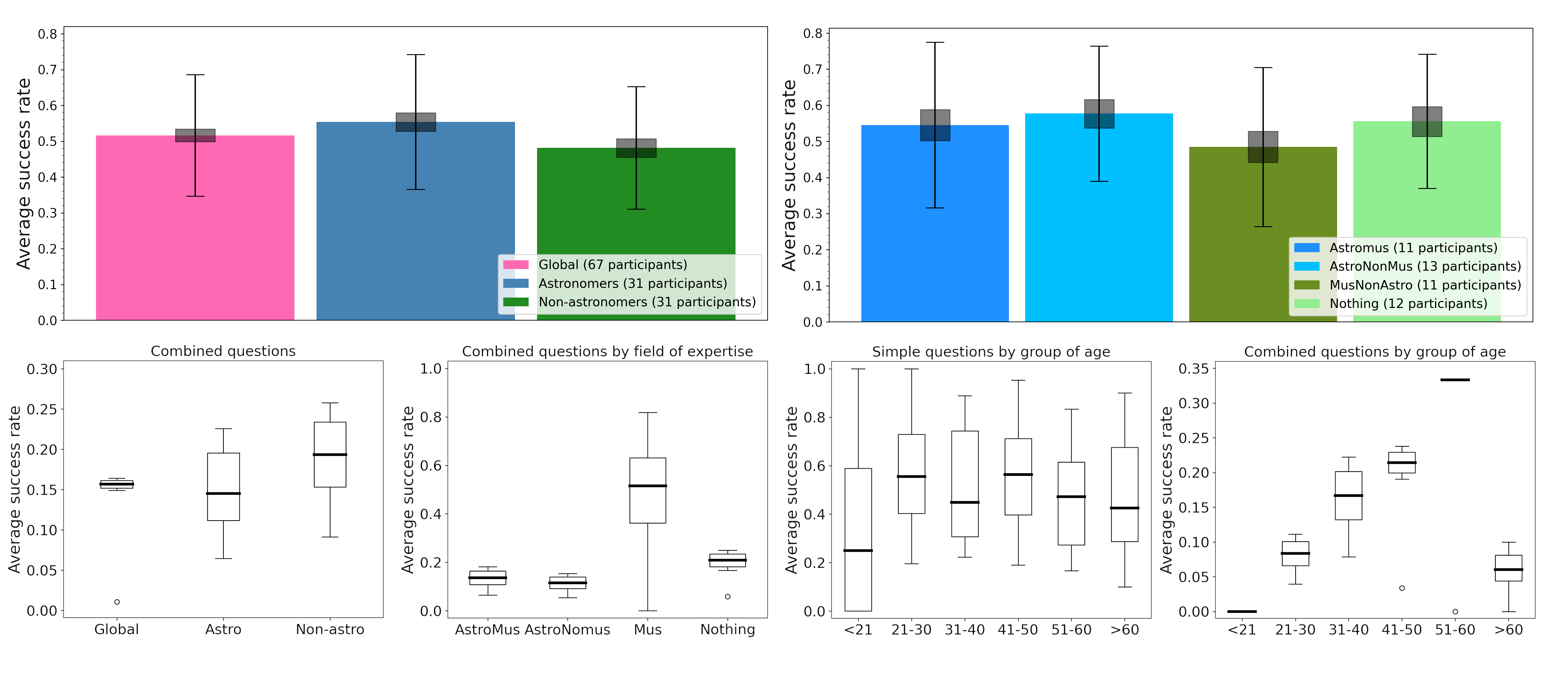}
    \caption{Evaluation results. \textbf{Up-left:} Average success rates for 67 participants on simple questions (left, magenta), for 31 professional astronomers (center, blue) and for 31 non-astronomers (right, green). \textbf{Up-right:} Average success rates on simple questions by field of expertise (balanced subgroups). From left to right: Astronomers musicians, Astronomers no musicians, Musicians no astronomers, and non-experienced in any of the fields. \textbf{Down-left:} Average success rates on combined questions, global and subgroup results. \textbf{Down-right:} Average success rates on simple and combined questions by age groups.}
    \label{fig:Results_figure}
\end{figure*}

\subsection{General results}
The survey was completed by 67 participants\footnote{The survey results and analysis notebooks are available at:

\url{\urlVCeval}}, including 31 professional astronomers, two of them identified as blind or low vision (BLV), and 36 non-astronomers. Their ages ranged from less than 21 (1) to more than 60 years old (10), with most of the participants ranging between 21 and 30 (18), and between 41 and 50 (21). They were mainly from Spain (50 participants) but also from Mexico, USA, Japan, Germany, China, Malta, Australia, and UK. Participants were asked about their music preferences and native language to explore whether language influences the ability to recognize sound features. Although the study included speakers of eleven different languages, the sample size was too small and diverse to draw any definitive conclusions. The same limitation applied to the analysis of music preferences.

As shown in the first graph of Fig.~\ref{fig:Results_figure} the mean global success rate obtained by 67 participants was 0.516, with a Jeffreys confidence interval of (0.498, 0.534), standard deviation 0.169, and 68.3\% of uncertainty. The subgroup formed by professional astronomers (31 participants) obtained a mean success rate of 0.554 with a Jeffreys confidence interval of (0.528, 0.579), standard deviation of 0.188, and the subgroup of non-astronomers (randomly down-sampled from 36 to 31 participants to allow direct comparison) obtained a mean success rate of 0.481 with a Jeffreys confidence interval of (0.455, 0.507), standard deviation 0.171, and 68.3\% of uncertainty. 

Although the sample was too small to establish statistical significance, these indicative results appear to confirm that all participants were able to understand the information from the sonifications thanks to the training videos, even without having previous experience in Astronomy. Agreeing with the results obtained using auditory graphs by~\cite{smith2005effects}, the training and context provided in the survey enhanced the performance of the participants. This is particularly notable in the analysis of the combined questions, in which non-astronomers performed 1.33 times better than professional astronomers, as can be noticed in the left-down graph of Fig.~\ref{fig:Results_figure}. This result could be related to the decision of allowing participants to repeat the training videos as many times as they wanted, that could benefit low-intermediate prior knowledge participants~\citep{van2022influence}.

As for the performance of BLV professional astronomers (only two participants), their results on simple questions were similar, although slightly higher (8\%), than those obtained by an equivalent random sampled group of non-BLV professional astronomers. This suggests that the application could help in bringing IFS analysis closer to BLV astronomers. In the combined questions, none of the two BLV astronomers succeeded in the Type/Age section. In the following, their results are included in the group of professional astronomers.


\subsection{Subgroup quantitative analysis}
To provide further analysis of the recorded feedback, the participants were divided into four subgroups according to their self-declared level of expertise in Astronomy and Music. The Astronomers musicians subgroup included professional astronomers also identified as professional or amateur musicians. A second group of astronomers no musicians was used to analyze the influence of sound analysis experience on potential experts in the proposed analysis tasks. The Musicians no astronomers group was formed by participants identified as professional and amateur musicians. Finally, the non-experienced group included the rest of the participants declaring no experience in any of the fields. Table~\ref{tab:Results_table} summarizes the results obtained on simple and combined question sections.

\begin{figure*}
	\includegraphics[width=\linewidth]{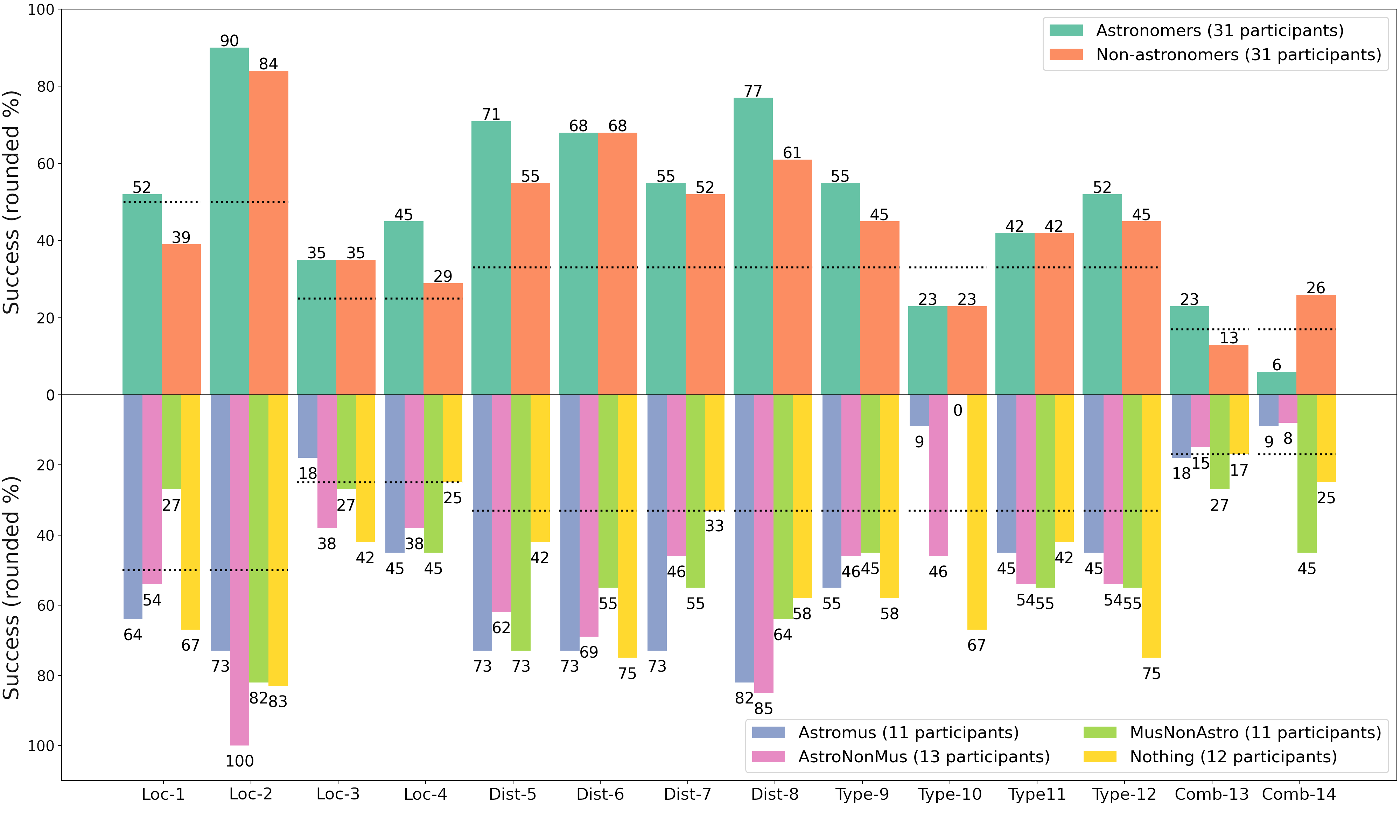}
    \caption{Success rate by question and group of expertise (notice the difference in the number of participants). Results for Astronomers vs Non-astronomers (up), and subgroup rates for Astronomers musicians, Astronomers non-musicians, Musicians non-astronomers, and Non-experienced participants (down). Sound location questions referenced as "Loc", distance to the center of the galaxy questions referenced as "Dist", type of galaxy questions referenced as "Type", and combined questions (success = all multiple choice options correct) referenced as "Comb". Dotted lines represent random-choice reference rate for each question. Notice random choice results for non-experienced participants in questions Loc-4, Dist-7, and Comb-13.}
    \label{fig:Questions_figure}
\end{figure*}

As shown in the up-right graph of Fig.~\ref{fig:Results_figure}, Astronomers no musicians were the best performers in the simple question sections, obtaining an average success rate of 0.577, with a Jeffreys confidence interval of (0.573, 0.616), standard deviation 0.187, and 68.3\% of uncertainty. It is worth mentioning that the non-experienced group performed as well as the Astronomer musicians, achieving an average success rate of approximately 0.55 with a standard deviation of 0.186, under the same conditions of uncertainty. This performance was 1.14 times better than that of the Musicians, which may suggest that the additional focus required to learn about unfamiliar fields helped the non-experienced participants with the proposed tasks.

The number of correct responses by group of expertise is provided in Fig.~\ref{fig:Questions_figure}. As for the combined questions, the success rates obtained were notably lower than those obtained in simple questions for all groups, with the exception of Musicians no astronomers. This suggests that the experience in the analysis of sound events was more helpful than previous knowledge of the data for the proposed task. 

\begin{figure}
	\includegraphics[width=\columnwidth]{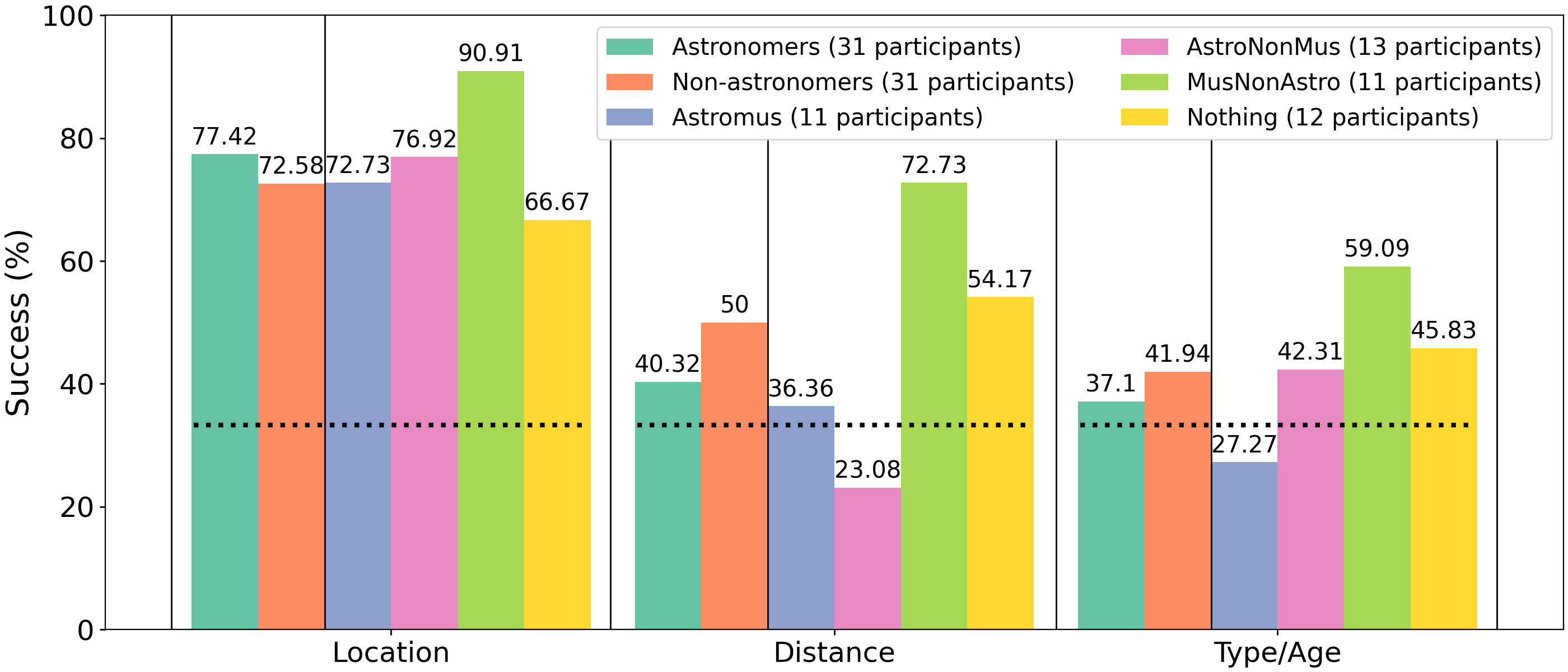}
    \caption{Average success rate of the two combined questions by blocks (Location, Distance, and Type/Age). From left to right, success rate for Astronomers vs Non-astronomers, and subgroup rates for Astronomers musicians, Astronomers non-musicians, Musicians non-astronomers, and Non-experienced participants. The dotted line indicates the success rate expected by random choice (0.33) considering three possible responses by block (correct, incorrect, and not answered), since 19 participants did not enter any response in some blocks.}
    \label{fig:Comb_figure}
\end{figure}

Further analysis of the combined responses revealed how participants failed to answer all three aspects of the combined questions simultaneously, but performed well per section. From the complete sample, 74.63\% of the participants located the sonification correctly, 46.27\% marked the correct distance to the center of the galaxy, and 36.57\% successfully interpreted the age of the galaxy. The respective averaged success rates from the simple questions per section were 53.2\% for the location questions, 63.30\% for distance analysis questions, and 40.7\% for the type of galaxy questions. 

As illustrated by Fig.~\ref{fig:Comb_figure}, the Type/Age questions had the lowest relative success rates across all groups, possibly due to the level of abstraction involved in these tasks. This result suggests that interpreting a galaxy's type or age through sound in this specific sonification implementation requires more training than the other tasks, which were more intuitive and aligned with the participants' prior experience. It is worth mentioning the exception of the astronomers no musicians, who obtained the worst results in the interpretation of the distance to the center of the galaxy. This fact has a positive correlation with the lack of accuracy in distance perception, when compared to horizontal localization, as discussed by \cite{middlebrooks2015sound}.

Analyzing the results of the combined questions by age (down-right graph of Fig.~\ref{fig:Results_figure}), the group ranging from 51 to 60 years old (6 participants) performed 1.56 times better than the 41-50 subgroup (21 participants), and 1.73 times better than the 31-40 subgroup (9 participants), suggesting that experience and attention to detail played an important role in the understanding of complex auditory information. Notice that the size of the samples do not provide statistical significance.

\begin{figure}
	\includegraphics[width=\columnwidth]{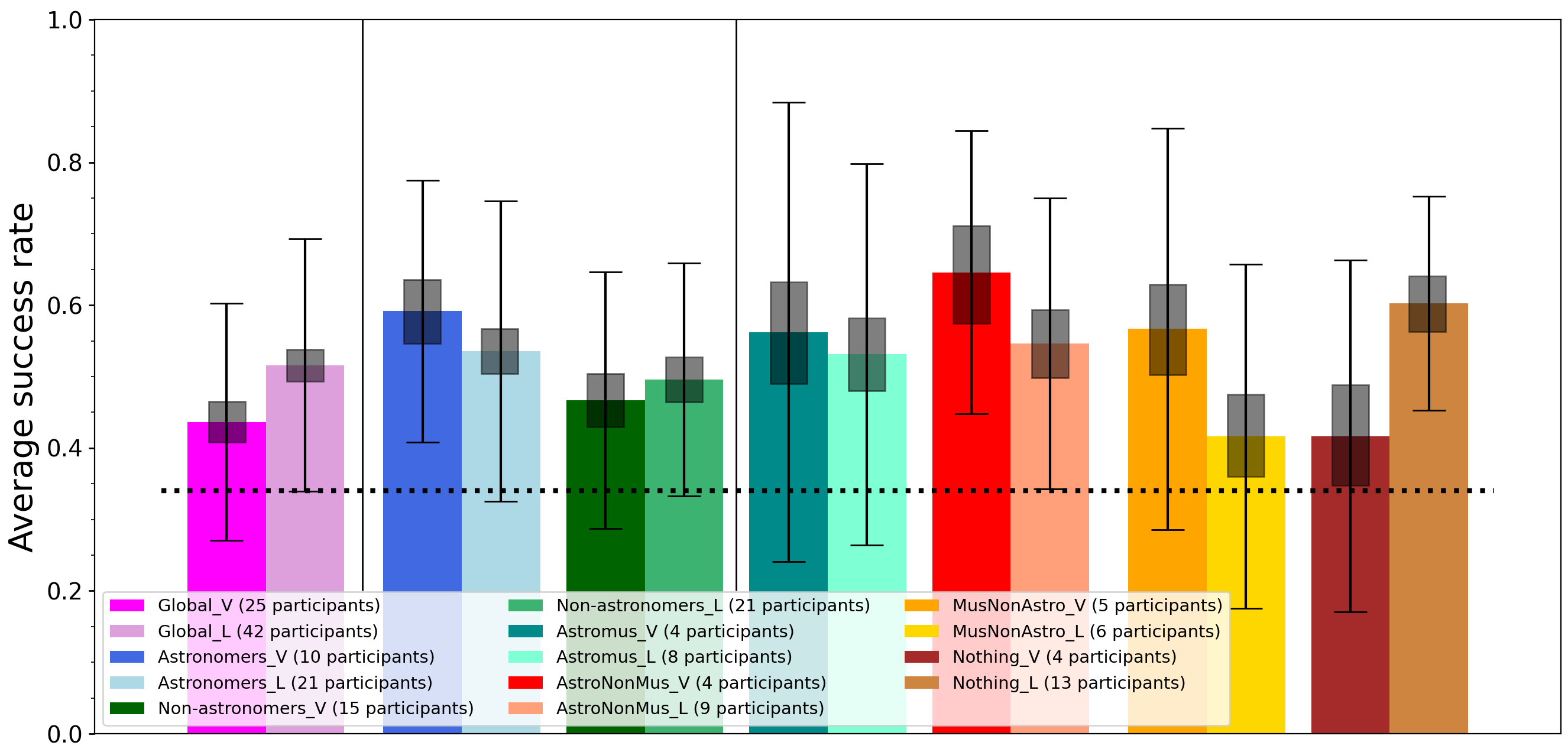}
    \caption{Success rate on simple questions for participants trained only with videos (V) vs participants testing the application live (L). From left to right, global results, astronomers vs non-astronomers, and expertise subgroups. Dotted line shows averaged random choice rate (0.34) from questions with 2, 3, and 4 possible responses.}
    \label{fig:VideoVsLive}
\end{figure}

Fig.~\ref{fig:VideoVsLive} provides an additional comparison between the results of the 42 participants that could try the application in person and the 25 participants that only used the training videos. Live testers presented a global success rate of 0.51 vs the 0.43 of only video trained participants. Nevertheless, the best results were obtained by astronomers non-musicians subgroup (4 participants) using only the training videos with no repetition restrictions (0.65), followed by non-experienced subgroup (13 participants) testing the application live (0.60). These results suggest that both training methods were useful for the proposed tasks.

\begin{table}
	\centering
	\caption{Quantitative evaluation. Results for simple and combined question sections shown by group of expertise and age. BLV astronomers were included respectively in AstroMus and AstNoMus groups}
	\label{tab:Results_table}
	\begin{tabular}{lccccc} 
		   & Answers & Success & std & Comb.succ & Comb.std\\
		\hline
		Global & 67 & 0.516 & 0.169 & 0.157 & 0.011\\
            \hline
		Astro & 31 & 0.554 & 0.188 & 0.145 & 0.114\\
		NoAstro & 31 &0.481 & 0.1709 & 0.193 & 0.091\\
		\hline
            AstroMus & 11 & 0.545 & 0.229 & 0.136 & 0.064\\
            AstNoMus & 13 & 0.577 & 0.187 & 0.115 & 0.054\\
            MusNoAst & 11 & 0.485 & 0.220 & 0.364 & 0.128\\
            Nothing & 12 & 0.555 & 0.186 & 0.208 & 0.059\\
            \hline
            <21 & 1 &0.250 & 0.452 & 0.0 & 0.0\\
            21-30 & 18 & 0.555 & 0.195 & 0.083 & 0.039\\
            31-40 & 9 & 0.509 & 0.229 & 0.166 & 0.078\\
            41-50 & 21 &0.555 & 0.193 & 0.214 & 0.033\\
            51-60 & 6 &0.444 & 0.217 & 0.333 & 0.0\\
            >60 & 10 &0.450 & 0.247 & 0.05 & 0.070\\
            \hline
            BLV & 2 & 0.542 & 0.396 & 0 & 0\\
            \hline
	\end{tabular}
\end{table}

\subsection{Qualitative Feedback}
\begin{figure}
	\includegraphics[width=\columnwidth]{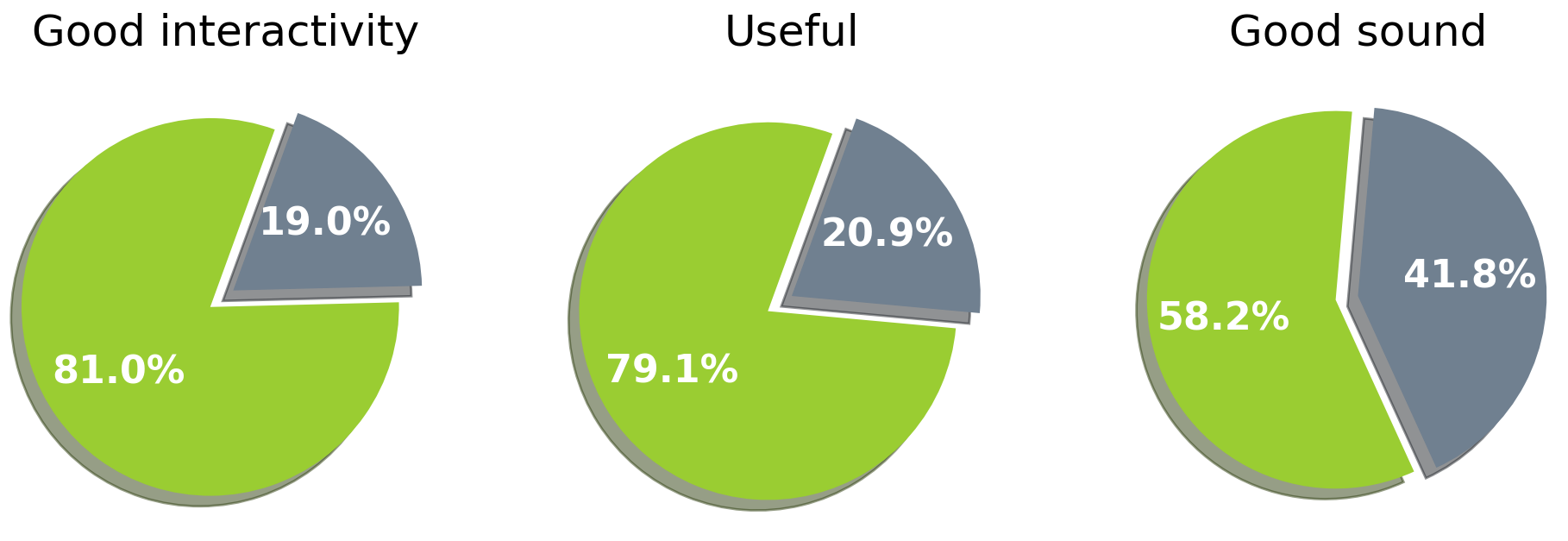}
    \caption{Qualitative evaluation. Interactivity: feedback from 42 participants who tested the application "in person". Usefulness and aesthetics: full sample, 67 participants. 81\% declared that the application had a "good interactivity", 79.1\% found it "useful" and 58.2\% "good sounding".}
    \label{fig:Qualitative_figure}
\end{figure}

To evaluate the interactivity, usefulness, and aesthetics of the proposal, the three qualitative questions described in Section~\ref{sec:Survey_design} were included in the survey. Of the 67 participants that completed the survey, 42 tested the application in person. As shown in Fig.~\ref{fig:Qualitative_figure}, 81\% of these participants expressed the good interactive response of the application, 79.1\% of the complete sample of participants (67) found the application "Useful" or "Very useful", 19.4\% "Doubtfully useful", and one participant (1.49\%) considered it "Usefulness". Regarding the aesthetics, 58.2\% rated it as "Good" or "Nice sounding", 34.33\% as "Acceptable", and 5.97\% as "Bad sounding". The subgroup results are also available in Table~\ref{tab:Qualitative_table}. 
Additionally, 7 participants explicitly expressed that they found it difficult to differentiate the sounds, and 8 participants explicitly expressed their enthusiasm about the project.

\begin{table}
	\centering
	\caption{Qualitative evaluation. Percentages by group of expertise. Asterisk values correspond to the sample of participants that tested the application in person. BLV astronomers were included respectively in AstroMus and AstNoMus groups}
	\label{tab:Qualitative_table}
	\begin{tabular}{lcccc} 
		   & Answers & Good Interactivity* & Useful & Good Sound\\
		\hline
		Global & 42*/67 & 80.95 & 79.10 & 58.21\\
            \hline
		Astro & 21*/31 & 76.19 & 74.19 & 51.61\\
		NoAstro & 17*/31 & 82.35 & 80.64 & 64.52\\
		\hline
            AstroMus & 8*/11 & 75.0 & 63.64 & 45.45\\
            AstNoMus & 9*/13 & 66.66 & 79.92 & 53.85\\
            MusNoAst & 6*/11 & 83.33 & 81.82 & 63.64\\
            Nothing & 9*/12 & 55.55 & 83.33 & 58.33\\
            \hline
            BLV & 1*/2 & 100.0 & 100.0 & 100.0\\
            \hline
            
	\end{tabular}
\end{table}

\section{Conclusions}

The design and evaluation of fast and efficient multimodal interactive tools for the exploration of IFS datacubes can help in the analysis of current massive spectroscopic surveys, complementing the possibilities of visual representations, and fostering inclusion and accessibility.

This article provides a summary of the motivations and design strategies used in the development of the interactive multimodal binaural application \textit{ViewCube}, and a user study with specialized and non-specialized participants analyzing selected galaxies from the CALIFA survey. The complete work was aimed at exploring the potential of multimodal IFS for the analysis of datacubes, with the motivation of making them more accessible for blind and low vision (BLV) researchers, and more immersive for complementary exploration through sound.
 
The tool allows the exploration of a wide variety of datacubes from different instruments and surveys across various wavelength ranges and includes a deep learning sonification approach to provide accurate comprehensible sonifications of the spectral information of the datacubes. The approach serves as an initial, general-purpose sonification tool designed to provide an overview of spectral properties, particularly in terms of stellar age and emission lines. In this context, the autoencoder effectively captures the general characteristics of both the strong emission lines (if present) and the continuum.

Regarding the qualitative feedback obtained from the 42 participants (including 21 professional astronomers) who evaluated the application in person, it can be concluded that the interactivity of the application is good or very good (80.95\%), with 79.1\% of the complete sample of participants (including 31 professional astronomers) finding it "Useful" or "Very useful", and 58.21\% declaring it as "Good" or "Nice sounding".

The quantitative evaluation of the application was conducted through an online survey featuring five training videos. The questionnaire was structured in four blocks aimed at analyzing the potential of the application for the estimation by sound of: 1) the position of a user-selected spaxel in the virtual soundscape generated by a datacube (left, right, front, or rear); 2) the distance of the user-selected spaxel to the center of the represented galaxy (close to the center, intermediate distance, or far from the center); 3) the type/age of the spectrum of the user-selected spaxel (star-forming region, intermediate age, or retired galaxy); and 4) all three characteristics combined.

Although the sample was too small to provide statistical significance, the results suggest that all participants (experienced and non-experienced) were able to retrieve information from the sonifications, presenting an average success rate of 0.516, with professional astronomers performing 1.15 times better than non-astronomers. 

The sample included two professional astronomers self-declared BLV. This group presented an average success rate of 0.541, 8\% higher than a random sampled subgroup formed by two professional astronomers, and 5\% higher than the mean of the complete sample, suggesting that the application can improve the access of BLV astronomers to IFS analysis. In the following analysis, BLV participants were included in professional astronomer subgroups.

The subgroup analysis revealed how the astronomers no musicians obtained the higher success rates, followed by astronomers musicians and non-experienced participants, who performed 1.14 times better than musicians. These results suggest that the additional attention required for learning aspects of unfamiliar fields could have helped the non-experienced participants with the proposed tasks.

Concerning the combined questions, the non-experienced group performed even slightly better than the astronomers (1.33 times). In these questions,  musicians were the top performers, suggesting that, for the proposed task, experience in analyzing sound events was more beneficial than prior knowledge of the data. Analyzing the results by age, participants between 51 and 60 years old were the top performers, suggesting that experience and attention to detail played a significant role in understanding complex auditory information.

In conclusion, although further research with additional sonification approaches and alternative spectroscopic surveys is planned, the promising trends outlined in this article suggest that the use of multimodal IFS displays can enhance the datacube analysis process and make 3D spectroscopy more accessible to BLV researchers. \textit{ViewCube} has demonstrated its ability to convey information about CALIFA’s galaxies, which was understood by both experienced and non-experienced users.

\section*{Acknowledgements}
We extend our gratitude to the anonymous referee for their valuable
and insightful feedback, which has contributed to enhancing the
quality and clarity of our manuscript.

We want to thank all the anonymous volunteers who made this research possible by testing the application and completing the survey.

This study uses data provided by the Calar Alto Legacy Integral Field Area (CALIFA) survey (\url{https://califa.caha.es/}). Based on observations collected at the Centro Astronómico Hispano Alemán (CAHA) at Calar Alto, operated jointly by the Max-Planck-Institut fűr Astronomie and the Instituto de Astrofísica de Andalucía (CSIC).

R.G.B. acknowledges financial support from the Severo Ochoa grant CEX2021-001131- S funded by MCIN/AEI/ 10.13039/501100011033 and PID2022-141755NB-I00. 

\section*{Data Availability}
The datacubes from the CALIFA Survey are available at \url{https://califa.caha.es/FTP-PUB/reduced/COMB/reduced_v2.2/}.

The encoded cubes generated for the sonification can be found at \url{https://zenodo.org/records/10570065}.

The feedback recorded from the survey and its analysis notebooks are available at \url{\urlVCeval}.



\bibliographystyle{rasti}
\bibliography{multimodal} 

\appendix

\section{Synthesizer description}
\label{sec:apendix}
The equation of the six-oscillator additive synthesizer implemented in \textit{SoniCube} can be expressed for each sonification as:
\begin{equation}
\label{eq:Synth}
S(t) = \sum_{i=0}^{6} A\textsubscript{i} F r \sin{(2\pi f\textsubscript{i} t + \phi\textsubscript{i})}     
\end{equation} 

where \textit{A\textsubscript{i}} is the A-weighting coefficient for each frequency, \textit{F} is the median of absolute flux of the represented galaxy spectrum, \textit{r} is the ratio or slope of the dynamic range limiter/compressor used to control salient flux values, \textit{f\textsubscript{i}} are the fundamental frequencies obtained through the autoencoder dimension reduction process (6 dimensions), and \textit{$\phi$\textsubscript{i}}, the relative phases of the oscillators (in our case, all set to zero).

Attending the loudness of each sonification, A-weighting coefficients are calculated for each one of the six frequencies or formants as~\citep{iec201361672}:
\begin{equation}
A\textsubscript{i} = 20\log{R\textsubscript{A}(f\textsubscript{i})}-20\log{R\textsubscript{A}(1000)} \approx 20\log{R\textsubscript{A}(f\textsubscript{i})} + 2
\end{equation} 

with \textit{R\textsubscript{A}(f\textsubscript{i})} calculated from the expression:

\begin{equation}
R\textsubscript{A}(f) = \frac{12194^2 f^4}{(f^2+20.6^2) \sqrt{(f^2+1007.7^2)(f^2+737.9^2)(f^2+12194^2)}}
\end{equation} 

\begin{figure}
	\includegraphics[width=\columnwidth]{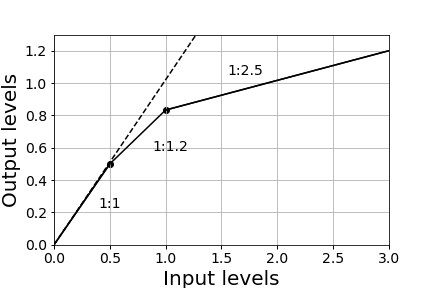}
    \caption{Transfer level curve used for the control of salient flux values that may cause ear damage when representing unexplored datacubes.}
    \label{fig:Compressor}
\end{figure}

Fig.\ref{fig:Compressor} shows the transfer level curve of the two-stage dynamic range limiter/compressor included in \textit{SoniCube} for preventing ear damage. The slope of each stage is represented by \textit{r} in equation~\ref{eq:Synth}.

A stereo reverberation processor based on eight delay lines~\citep{kahrs1998applications} is added to the signal flow to provide information about the distance from the selected spaxel to the center of the galaxy. The final signal sent to the binaural encoder can be expressed as:

\begin{equation}
X\textsubscript{L,R}(t) = S(t) +  d \sum_{j=0}^{8} g\textsuperscript{j-1} S(t-j \tau)
\end{equation} 

where \textit{S(t)} is the output of the additive synthesizer, \textit{d} is the distance of the user-selected spaxel to the reference point (center of the galaxy), \textit{g} is the gain of the feedback loop (in our case set to .9 to provide a long reverberation effect), and \textit{$\tau$} is the fixed delay time applied to each line.

\section{Binaural encoding}
\label{sec:apendix2}

Human hearing can locate sound sources in a 3D space through the analysis of interaural level differences (ILD), and interaural time delays (ITD).
The alterations produced in a sound when traveling from the source to the listener can be characterized using Head Related Transfer Functions (HRTF) \citep{brown1997efficient}. Binaural systems are based on the convolution of the HRTF of both ears with the sound sources, allowing the spatialization of static and dynamic locations~\citep{lazzarini2008new}. 

The relationship between the sound source and the signal reaching the listener ears can be expressed in terms of the azimuth (AZ) and elevation (EL) angles, the distance to the source (d), and the angular frequency (w), as shown in the following equation:
\begin{equation}
Y\textsubscript{L,R}(AZ, EL, d, w) = H\textsubscript{L,R}(AZ, EL, d, w)* X\textsubscript{L,R}(w)       
\end{equation} 
where Y\textsubscript{L,R} are the audio spectra of acoustic signals at listener's ears, H\textsubscript{L,R} is the HRTF and X\textsubscript{L,R} are the audio spectra of the sound source. 

\textit{Sonicube} uses the binaural \textit{hrtfmove} \textit{Csound} opcode~\citep{ref_hrtfmove2} to represent the spectral information of the datacubes in an immersive 2D soundscape correlated to \textit{ViewCube}'s UI.
The azimuth is calculated from the (x,y) coordinates and the elevation is set to zero to discard the third dimension. Distance is not used to keep the levels of the sonification flux-dependent, reducing the general expression to: 
\begin{equation}
Y\textsubscript{L,R}(AZ, 0, w) = H\textsubscript{L,R}(AZ, 0, w) * X\textsubscript{L,R}(w)       
\end{equation}

\bsp	
\label{lastpage}
\end{document}